\documentclass[fleqn,a4paper]{article}

\usepackage[utf8]{inputenc} % why not type "Bézout" with unicode?
\usepackage[T1]{fontenc} % vector fonts plz
\usepackage{fullpage,amsmath,amssymb,latexsym,graphicx}
\usepackage{newtxtext,newtxmath} % Times for PR
\usepackage{appendix}
\usepackage[dvipsnames]{xcolor}
\usepackage[
  colorlinks=true,
  urlcolor=MidnightBlue,
  citecolor=MidnightBlue,
  filecolor=MidnightBlue,
  linkcolor=MidnightBlue
]{hyperref} % ref and cite links with pretty colors
\usepackage[
  style=phys,
  eprint=true,
  maxnames = 100
]{biblatex}
\usepackage{anyfontsize,authblk}

\begin{document}

\title{
  When is the average number of saddle points typical?
}

\author{Jaron Kent-Dobias}
\affil{Istituto Nazionale di Fisica Nucleare, Sezione di Roma I}

\maketitle
\begin{abstract}
  A common measure of a function's complexity is the count of its stationary
  points. For complicated functions, this count grows exponentially with the
  volume and dimension of their domain. In practice, the count is averaged over
  a class of functions (the annealed average), but the large numbers
  involved can produce averages biased by extremely rare samples. Typical
  counts are reliably found by taking the average of the logarithm (the
  quenched average), which is more difficult and not often done in practice.
  When most stationary points are uncorrelated with each other, quenched and
  anneals averages are equal. Equilibrium heuristics can guarantee when most of
  the lowest minima will be uncorrelated. We show that these equilibrium
  heuristics cannot be used to draw conclusions about other minima and saddles
  by producing examples among Gaussian-correlated functions on the hypersphere
  where the count of certain saddles and minima has different quenched and
  annealed averages, despite being guaranteed `safe' in the equilibrium
  setting. We determine conditions for the emergence of nontrivial correlations
  between saddles, and discuss the implications for the geometry of those
  functions and what out-of-equilibrium settings might be affected.
\end{abstract}

Random high-dimensional energies, cost functions, and interaction networks are
important across disciplines: the energy landscape of glasses, the likelihood
landscape of machine learning and inference, and the interactions between
organisms in an ecosystem are just a few examples \cite{Stein_1995_Broken, Krzakala_2007_Landscape, Altieri_2021_Properties, Yang_2023_Stochastic}. A traditional tool for
making sense of their behavior is to analyze the statistics of points where
their dynamics are stationary \cite{Cavagna_1998_Stationary,
Fyodorov_2004_Complexity, Fyodorov_2007_Density, Bray_2007_Statistics}. For
energy or cost landscapes, these correspond to the minima, maxima, and saddles,
while for ecosystems and other non-gradient dynamical systems these correspond
to equilibria of the dynamics. When many stationary points are present, the
system is considered complex.

Despite the importance of stationary point statistics for understanding complex
behavior, they are often calculated using an uncontrolled approximation.
Because their number is so large, it cannot be reliably averaged. The annealed
approximation takes this average anyway, risking a systematic bias by rare and
atypical samples. The annealed approximation is known to be exact for certain
models and in certain circumstances, but it is used outside those circumstances
without much reflection \cite{Wainrib_2013_Topological, Kent-Dobias_2021_Complex,
Gershenzon_2023_On-Site}. In a few cases researchers have instead made the
better-controlled quenched average, which averages the logarithm of the number
of stationary points, and find deviations from the annealed approximation with
important implications for behavior \cite{Muller_2006_Marginal,
Ros_2019_Complex, Kent-Dobias_2023_How, Ros_2023_Quenched, Ros_2023_Generalized}. Generically,
the annealed approximation to the complexity is wrong when a nonvanishing
fraction of pairs of stationary points have nontrivial correlations in their
mutual position.

A heuristic line of reasoning for the appropriateness of the annealed
approximation is sometimes made when the approximation is correct for an
equilibrium calculation on the same system. The argument goes like this: since
the limit of zero temperature in an equilibrium calculation
concentrates the Boltzmann measure onto the lowest set of minima, the equilibrium free
energy in the limit to zero temperature will be governed by the same
statistics as the count of that lowest set of minima. This argument is strictly
valid only for the lowest minima, which at least in glassy problems are
rarely relevant to dynamical behavior. What about the \emph{rest} of the
stationary points?

In this paper, we show that the behavior of the ground state, or \emph{any}
equilibrium behavior, does not govern whether stationary points will have a
correct annealed average. In a prototypical family of models of random
functions, we determine a condition for when annealed averages
should fail and some stationary points will have nontrivial correlations in their
mutual position. We produce examples of models whose equilibrium is guaranteed
to never see such correlations between thermodynamic states, but where a
population of saddle points is nevertheless correlated.

We study the mixed spherical models, which are models of Gaussian-correlated
random functions with isotropic statistics on the $(N-1)$-sphere. Each model
consists of a class of functions $H:S^{N-1}\to\mathbb R$ defined by the
covariance between the functions evaluated at two different points
$\pmb\sigma_1,\pmb\sigma_2\in S^{N-1}$, which is a function of the scalar
product (or overlap) between the two configurations:
\begin{equation} \label{eq:covariance}
  \overline{H(\pmb\sigma_1)H(\pmb\sigma_2)}=\frac1Nf\bigg(\frac{\pmb\sigma_1\cdot\pmb\sigma_2}N\bigg)
\end{equation}
Specifying the covariance function $f$ uniquely specifies the model. The series
coefficients of $f$ need to be nonnnegative in order for $f$ to be a
well-defined covariance. The case where $f$ is a homogeneous polynomial has
been extensively studied, and corresponds to the pure spherical models of glass
physics or the spiked tensor models of statistical inference \cite{Castellani_2005_Spin-glass}. Here we will
study cases where $f(q)=\frac12\big(\lambda q^3+(1-\lambda)q^s\big)$ for
$\lambda\in(0,1)$, called $3+s$ models. These are examples of \emph{mixed}
spherical models, which have been studied in the physics and statistics
literature and host a zoo of complex orders and phase transitions
\cite{Crisanti_2004_Spherical, Crisanti_2006_Spherical,
Krakoviack_2007_Comment, Crisanti_2007_Amorphous-amorphous,
Crisanti_2011_Statistical, BenArous_2019_Geometry, Subag_2020_Following, ElAlaoui_2020_Algorithmic}.

\begin{figure}
  \centering
  \includegraphics{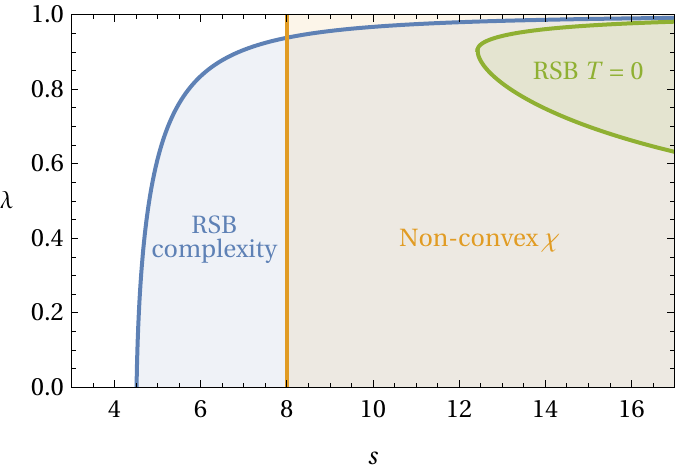}
  \caption{
    A phase diagram of the boundaries we discuss in this paper for the $3+s$
    model with $f=\frac12\big(\lambda q^3+(1-\lambda)q^s\big)$. The blue region
    shows models which have some stationary points with nontrivial correlated
    (\textsc{rsb}) structure, and is given by $G_f>0$ where $G_f$ is found in
    \eqref{eq:condition}. The yellow region shows where $\chi(q)=f''(q)^{-1/2}$
    is not convex and therefore nontrivial correlations between states are
    possible in equilibrium. The green region shows where nontrivial
    correlations exist at the ground state, adapted from
    \cite{Auffinger_2022_The}. We find that models where correlations between
    equilibrium states are forbidden can nonetheless harbor
    correlated stationary points.
  } \label{fig:phases}
\end{figure}

There are several well-established results on the equilibrium of this model.
First, if the function $\chi(q)=f''(q)^{-1/2}$ is convex then it is not possible for the
equilibrium solution to have nontrivial correlations between states at any
temperature \cite{Crisanti_1992_The}.\footnote{
  More specifically, convex $\chi$ cannot have an equilibrium order with more than
  {\oldstylenums1\textsc{rsb}} order among the configurations. In equilibrium,
  {\oldstylenums1\textsc{rsb}} corresponds to trivial correlations between
  thermodynamic states, but nontrivial correlations exist \emph{within} a state
  at nonzero temperature. When temperature goes to zero,
  {\oldstylenums1\textsc{rsb}} in equilibrium reduces to replica symmetry among
  the lowest-lying states. Because in this paper we focus on symmetry breaking
  between stationary points, we consider this form of \textsc{rsb} in
  equilibrium trivial because it does not imply any nontrivial correlations
  between states.
}
This is a strong condition on the form of equilibrium order. Note that
non-convex $\chi$ does not imply that you \emph{will} see nontrivial correlations between
states at some temperature. In the $3+s$ models we consider here, models with
$s>8$ have non-convex $\chi$ and those with $s\leq8$ have convex $\chi$ independent
of $\lambda$. Second, the characterization of the ground state has been made
\cite{Crisanti_2004_Spherical, Crisanti_2006_Spherical,
Crisanti_2011_Statistical, Auffinger_2022_The}. In the $3+s$ models we
consider, for $s>12.430...$ nontrivial ground state configurations appear in
a range of $\lambda$. These bounds on equilibrium order are shown in
Fig.~\ref{fig:phases}, along with our result for where the complexity has
nontrivial correlations between some stationary points. As evidenced in that
figure, correlations among saddles are possible well inside regions that
forbid them among equilibrium states.

There are two important features which differentiate stationary points
$\pmb\sigma^*$ in the spherical models: their \emph{energy density}
$E=\frac1NH(\pmb\sigma^*)$ and their \emph{stability}
$\mu=\frac1N\operatorname{\mathrm{Tr}}\operatorname{\mathrm{Hess}}H(\pmb\sigma^*)$.
The energy density gives the `height' in the landscape, while the
stability governs the spectrum of the stationary point.
In each spherical model, the spectrum of every stationary point is a Wigner
semicircle of the same width $\mu_\mathrm m=\sqrt{4f''(1)}$, but shifted by
constant. The stability $\mu$ sets this constant shift. When $\mu<\mu_\mathrm
m$, the spectrum has support over zero and we have saddles with an
extensive number of downward directions. When $\mu>\mu_\mathrm m$ the spectrum
has support only over positive eigenvalues, and we have stable minima.\footnote{
  Saddle points with a subextensive number of downward directions also exist
  via large deviations of some number of eigenvalues from the average spectrum.
} When $\mu=\mu_\mathrm m$, the spectrum has a pseudogap, and we have marginal minima.

The number $\mathcal N(E,\mu)$ of stationary points with energy density $E$ and
stability $\mu$ is exponential in $N$. Their complexity
$\Sigma(E,\mu)$ is defined by the average of the logarithm of their number:
$\Sigma(E,\mu)=\frac1N\overline{\log\mathcal N(E,\mu)}$. More often the
annealed complexity is calculated, where the average is taken before the
logarithm: $\Sigma_\mathrm a(E,\mu)=\frac1N\log\overline{\mathcal N(E,\mu)}$.
The annealed complexity has been computed for these models
\cite{BenArous_2019_Geometry, Folena_2020_Rethinking}, and the quenched
complexity has been computed for a couple examples which have nontrivial ground
states \cite{Kent-Dobias_2023_How}. The annealed complexity bounds the
complexity from above. A positive complexity indicates the presence of an
exponentially large number of stationary points of the indicated kind, while a
negative one means it is vanishingly unlikely they will appear. The line of
zero complexity is significant as the transition between many stationary points
and none.

In these models, trivial correlations between stationary points correspond with
zero overlap: almost all stationary points are orthogonal to each other. This
corresponds with \emph{replica symmetric} (\textsc{rs}) order. The emergence of
nontrivial correlations, and the invalidity of the annealed approximation, occurs
when some non-vanishing fraction of stationary point pairs have a nonzero
overlap. This corresponds to some kind of \emph{replica symmetry breaking}
(\textsc{rsb}). Here we restrict ourselves to a {\oldstylenums1}\textsc{rsb}
ansatz, which corresponds to two kinds of pairs of stationary point: a fraction
$x$ of pairs have the trivial zero overlap, and the remaining fraction $1-x$
have a nontrivial overlap $q_1$. In the annealed or replica-symmetric case,
$x=1$ and all but a vanishing fraction of stationary points are uncorrelated
with each other. Since other kinds of \textsc{rsb} order encompass
{\oldstylenums1}\textsc{rsb}, we are guaranteed that
$\Sigma\leq\Sigma_{\oldstylenums1\textsc{rsb}}\leq\Sigma_\mathrm a$. We will
discuss later in what settings the {\oldstylenums1}\textsc{rsb} complexity is
correct.

When the complexity is calculated using the Kac--Rice formula and a physicists'
tool set, the problem is reduced to the evaluation of an integral by the saddle
point method for large $N$ \cite{Kent-Dobias_2023_How}. The complexity is given
by extremizing an effective action,
\begin{equation}
  \Sigma_{\oldstylenums1\textsc{rsb}}(E,\mu)
  =\lim_{n\to0}\int dq_1\,dx\,\mathcal S_{\oldstylenums1\textsc{rsb}}(q_1,x\mid E,\mu)e^{nN\mathcal S_{\oldstylenums1\textsc{rsb}}(q_1,x\mid E,\mu)}
  =\mathop{\mathrm{extremum}}_{q_1,x}\mathcal S_{\oldstylenums1\textsc{rsb}}(q_1,x\mid E,\mu)
\end{equation}
for the action $\mathcal S_{\oldstylenums1\textsc{rsb}}$ given by
\begin{equation}
  \begin{aligned}
    &\mathcal S_{\oldstylenums1\textsc{rsb}}(q_1,x\mid E,\mu)
    =\mathcal D(\mu)
    +\mathop{\textrm{extremum}}_{\hat\beta,r_\mathrm d,r_1,d_\mathrm d,d_1}
    \Bigg\{
      \hat\beta E-r_\mathrm d\mu\\
      &\quad+\frac12\bigg[
        \hat\beta^2\big[f(1)-\Delta xf(q_1)\big]
        +(2\hat\beta r_\mathrm d-d_\mathrm d)f'(1)
        -\Delta x(2\hat\beta r_1-d_1)f'(q_1)
        +r_\mathrm d^2f''(1)-\Delta x\,r_1^2f''(q_1) \\
        &\quad+\frac1x\log\Big(
          \big(r_\mathrm d-\Delta x\,r_1\big)^2+d_\mathrm d\big(1-\Delta x\,q_1\big)-\Delta x\,d_1\big(1-\Delta xq_1\big)
          \Big)
          -\frac{\Delta x}x\log\Big(
            (r_\mathrm d-r_1)^2+(d_\mathrm d-d_1)(1-q_1)
          \Big)
      \bigg]
    \Bigg\}
  \end{aligned}
\end{equation}
where $\Delta x=1-x$ and
\begin{equation}
  \mathcal D(\mu)
  =\begin{cases}
    \frac12+\log\left(\frac12\mu_\text m\right)+\frac{\mu^2}{\mu_\text m^2}
     & \mu^2\leq\mu_\text m^2 \\
    \frac12+\log\left(\frac12\mu_\text m\right)+\frac{\mu^2}{\mu_\text m^2}
    -\left|\frac{\mu}{\mu_\text m}\right|\sqrt{\big(\frac\mu{\mu_\text m}\big)^2-1}
    -\log\left(\left|\frac{\mu}{\mu_\text m}\right|-\sqrt{\big(\frac\mu{\mu_\text m}\big)^2-1}\right) & \mu^2>\mu_\text m^2
  \end{cases}
\end{equation}
The details of the derivation of these expressions can be found in \cite{Kent-Dobias_2023_How}.
The extremal problem in $\hat\beta$, $r_\mathrm d$, $r_1$, $d_\mathrm d$, and
$d_1$ has a unique solution and can be found explicitly, but the resulting
formula is unwieldy. The action can have multiple extrema, but the one for which the complexity is
\emph{smallest} gives the correct solution. There is always a solution for
$x=1$ which is independent of $q_1$, corresponding to the replica symmetric
case, and with $\Sigma_\mathrm
a(E,\mu)=\mathcal S_{\oldstylenums1\textsc{rsb}}(E,\mu\mid q_1,1)$. The crux of
this paper will be to determine when this solution is not the global one.

It isn't accurate to say that a solution to the saddle point equations is
`stable' or `unstable.' The problem of solving the complexity in this way is
not a variational problem, so there is nothing to be maximized or minimized,
and in general even global solutions are not even local minima of the action.
However, the stability of the action can still tell us something about the
emergence of new solutions: when a new solution bifurcates from an existing
one, the action will have a flat direction. Unfortunately this is difficult to
search out, since one must know the parameters of the new solution, and $q_1$
is unconstrained and can take any value in the old solution.

\begin{figure}
  \centering
  \includegraphics{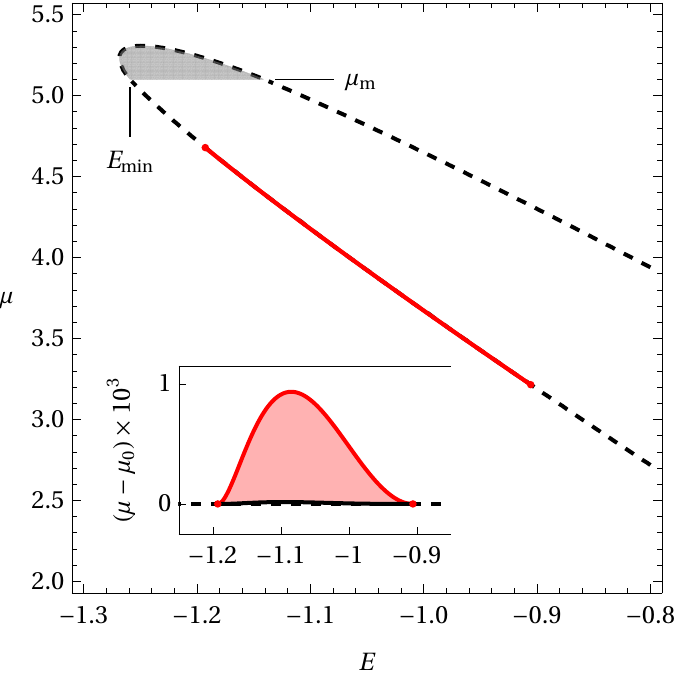}

  \caption{
    Stationary point statistics as a function of energy density $E$ and
    stability $\mu$ for a model with $f(q)=\frac12(\frac12q^3+\frac12q^5)$. The dashed black
    line shows the line of zero annealed complexity and
    enclosed inside the annealed complexity is positive. The solid black line (only visible in the inset) gives the line of zero {\oldstylenums1\textsc{rsb}} complexity. The red region (blown
    up in the inset) shows where the annealed complexity gives the wrong count
    and a {\oldstylenums1}\textsc{rsb} complexity in necessary. The red points
    show where $\det M=0$. The left point, which is only an upper bound on the
    transition, coincides with it in this case. The gray shaded region
    highlights the minima, which are stationary points with $\mu\geq\mu_\mathrm
    m$. $E_\textrm{min}$ is marked on the plot as the lowest energy at which
    extensive saddles are found.
  } \label{fig:complexity_35}
\end{figure}

There is one place where we can consistently search for a bifurcating solution
to the saddle point equations: along the zero complexity line $\Sigma_\mathrm
a(E,\mu)=0$. Going along this line in the replica symmetric solution, the
{\oldstylenums1}\textsc{rsb} complexity transitions at a critical point where
$x=q_1=1$ \cite{Kent-Dobias_2023_How}. Since all the parameters in the
bifurcating solution are known at this point, we can search for it by looking
for a flat direction. In the annealed solution for
points describing saddles ($\mu<\mu_\mathrm m$), this line is
\begin{equation} \label{eq:extremal.line}
  \mu_0=-\frac1{z_f}\left(2Ef'f''+\sqrt{2f''u_f\bigg(\log\frac{f''}{f'}z_f-E^2(f''-f')\bigg)}\right)
\end{equation}
where we have chosen the lower branch as a convention (see
Fig.~\ref{fig:complexity_35}) and where we define for brevity (here and
elsewhere) the constants
\begin{align}
  u_f&=f(f'+f'')-f'^2
  &&
  v_f=f'(f''+f''')-f''^2 \\
  w_f&=2f''(f''-f')+f'f'''
  &&
  y_f=f'(f'-f)+f''f
  &&
  z_f=f(f''-f')+f'^2
\end{align}
When $f$ and its derivatives appear without an argument, the implied argument is always 1, so, e.g., $f'\equiv f'(1)$.
If $f$ has at least two nonzero coefficients at second order or higher, all of
these constants are positive. Though in figures we focus on the lower branch of
saddles, another set of identical solutions always exists for $(E,\mu)\mapsto(-E,-\mu)$.
We also define $E_\textrm{min}$, the minimum energy at which saddle points with
an extensive number of downward directions are found, as the energy for which
$\mu_0(E_\mathrm{min})=\mu_\mathrm m$.

Let $M$ be the matrix of double partial derivatives of the action with
respect to $q_1$ and $x$. We evaluate $M$ at the replica symmetric saddle point
$x=1$ with the additional constraint that $q_1=1$ and along the extremal
complexity line \eqref{eq:extremal.line}. We determine when a zero eigenvalue
appears, indicating the presence of a bifurcating {\oldstylenums1}\textsc{rsb}
solution, by solving $0=\det M$. We find
\begin{equation}
  \det M
  =-\bigg(\frac{\partial^2\mathcal S_{\oldstylenums1\textsc{rsb}}}{\partial q_1\partial x}\bigg|_{\substack{x=1\\q_1=1}}\bigg)^2
  \propto(ay^2+bE^2+2cyE-d)^2
\end{equation}
where $y=-\frac12z_f\mu-f'f''E$ is proportional to the square-root term in
\eqref{eq:extremal.line} and the constants $a$, $b$, $c$, and $d$ are defined
by
\begin{equation}
  a=\frac{w_f\big(3y_f^2-4ff'f''(f'-f)\big)-6y_f^2(f''-f')f''}{(u_fz_ff'')^2f'}
  \qquad
  b=\frac{f'w_f}{z_f^2}
  \qquad
  c=\frac{w_f}{f''z_f^2}
  \qquad
  d=\frac{w_f}{f'f''}
\end{equation}
Changing variables from $\mu$ to $y$ is convenient because the branch
of \eqref{eq:extremal.line} is chosen by the sign of $y$ (the lower-energy
branch we are interested in corresponds with $y>0$). The relationship
between $y$ and $E$ on the extremal line is $g=2hy^2+eE^2$, where the constants
$e$, $g$, and $h$ are given by
\begin{equation}
  e=f''-f'
  \qquad
  g=z_f\log\frac{f''}{f'}
  \qquad
  h=\frac1{f''u_f}
\end{equation}

The solutions for $\det M=0$ can be calculated explicitly and correspond to
energies that satisfy
\begin{equation} \label{eq:energies}
  E_{\oldstylenums1\textsc{rsb}}^\pm
  =\operatorname{sign}(bg-de)\frac{-cg\pm\sqrt{c^2g^2+(2dh-ag)(bg-de)}}
  {
    \sqrt{2c^2eg+(2bh-ae)(bg-de)\mp2ce\sqrt{c^2g^2+(2dh-ag)(bg-de)}}
  }
\end{equation}
This predicts two points where a {\oldstylenums1}\textsc{rsb} solution can
bifurcate from the annealed one. The remainder of the transition line can be
found by solving the extremal problem for the action very close to one
of these solutions, and then taking small steps in the parameters $E$ and $\mu$
until it terminates. In many cases considered here, the line of transitions in
the complexity that begins at $E_{\oldstylenums1\textsc{rsb}}^+$, the higher
energy point, ends exactly at $E_{\oldstylenums1\textsc{rsb}}^-$, the lower
energy point, so that these two points give the precise range of energies at
which \textsc{rsb} saddles are found. An example that conforms with this
picture for a $3+5$ mixed model is shown in Fig.~\ref{fig:complexity_35}.

The expression inside the inner square root of \eqref{eq:energies} is
proportional to
\begin{equation} \label{eq:condition}
  G_f
  =
  f'\log\frac{f''}{f'}\big[
    3y_f(f''-f')f'''-2(f'-2f)f''w_f
  \big]
  -2(f''-f')u_fw_f
  -2\log^2\frac{f''}{f'}f'^2f''v_f
\end{equation}
If $G_f>0$, then the bifurcating solutions exist, and there are some saddles whose
complexity is corrected by a {\oldstylenums1\textsc{rsb}} solution.
Therefore, $G_f>0$ is a condition to see {\oldstylenums1}\textsc{rsb} in the
complexity. If $G_f<0$, then there is nowhere along the extremal line where
saddles can be described by such a complexity. The range of $3+s$ models where
$G_f$ is positive is shown in Fig.~\ref{fig:phases}.

\begin{figure}
  \centering
  \includegraphics{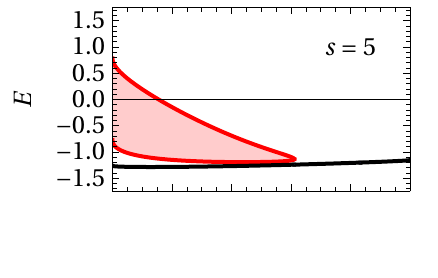}
  \hspace{-3em}
  \includegraphics{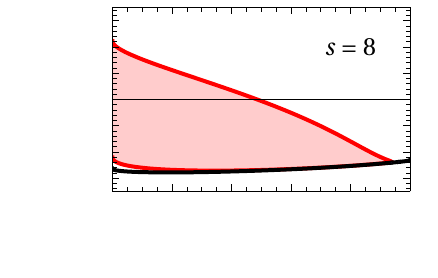}
  \hspace{-3em}
  \includegraphics{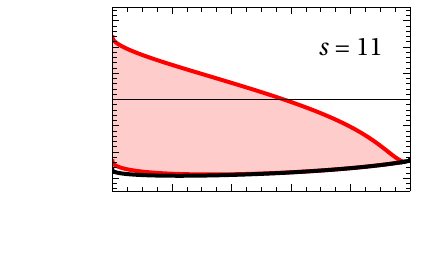}
  \hspace{-3em}
  \includegraphics{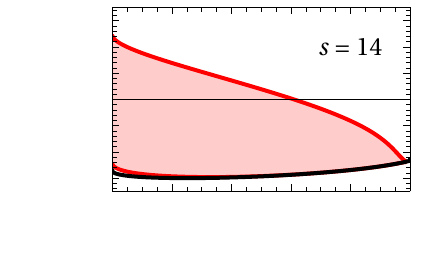} \\
  \vspace{-2em}
  \includegraphics{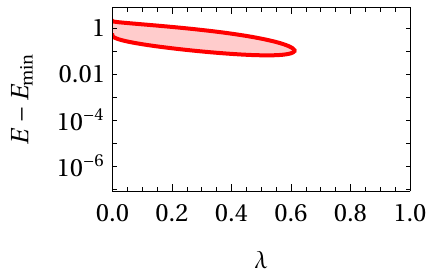}
  \hspace{-3em}
  \includegraphics{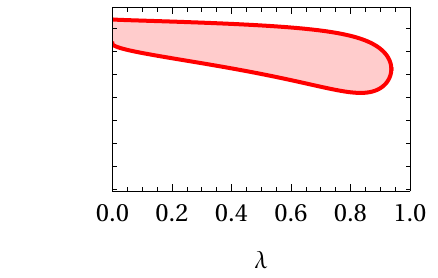}
  \hspace{-3em}
  \includegraphics{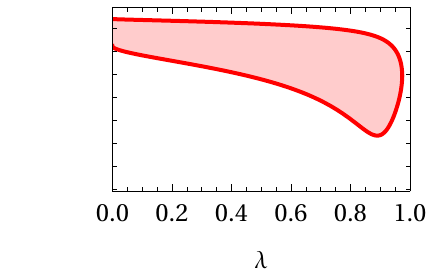}
  \hspace{-3em}
  \includegraphics{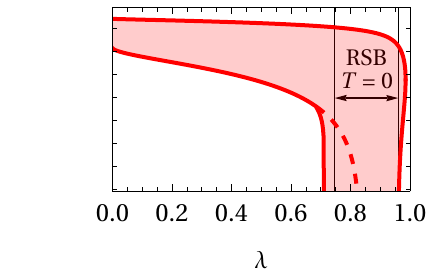}

  \caption{
    The range of energies where \textsc{rsb} saddles are found for the $3+s$
    model with varying $s$ and $\lambda$. In the top row the black line shows
    $E_\textrm{min}$, the minimum energy where saddles are found, and in the
    bottom row this energy is subtracted away to emphasize when the
    \textsc{rsb} region crosses into minima. For most $s$, both the top and
    bottom lines are given by $E_{\oldstylenums1\textsc{rsb}}^\pm$, but for $s=14$
    there is a portion where the low-energy boundary has $q_1<1$. In that plot,
    the continuation of the $E_{\oldstylenums1\textsc{rsb}}^-$ line is shown
    dashed. Also marked is the range of $\lambda$ for which the ground state
    minima are characterized by nontrivial \textsc{rsb}.
  } \label{fig:energy_ranges}
\end{figure}

\begin{figure}
  \centering
  \includegraphics{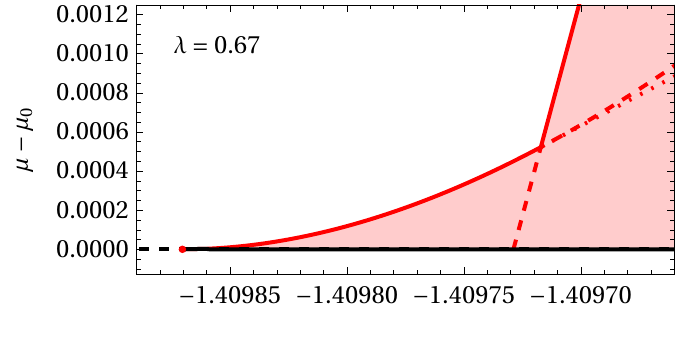}\\
  \vspace{-1em}
  \includegraphics{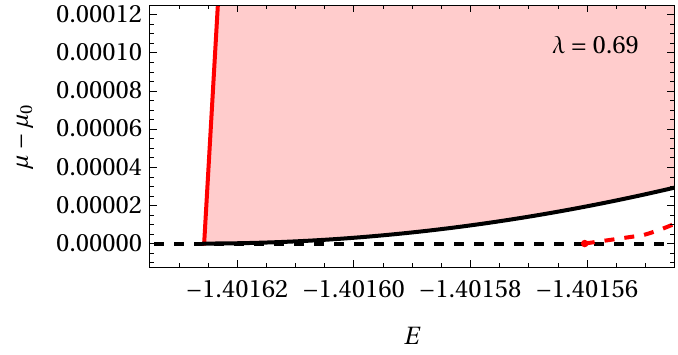}

  \caption{
    Examples of $3+14$ models where the solution
    $E_{\oldstylenums1\textsc{rsb}}^-$ does and doesn't define the lower limit
    of energies where \textsc{rsb} saddles are found. In both plots the red dot
    shows $E_{\oldstylenums1\textsc{rsb}}^-$, while the solid red lines shows
    the transition boundary with the \textsc{rs} complexity. The dashed black
    line shows the \textsc{rs} zero complexity line, while the solid black line
    shows the {\oldstylenums1}\textsc{rsb} zero complexity line. The dashed red
    lines show where a nonphysical {\oldstylenums1\textsc{rsb}} phase appears
    (the spinodal of that phase). The dotted red line shows an abrupt phase
    transition between different {\oldstylenums1}\textsc{rsb} phases.
    \textbf{Top:} $\lambda=0.67$. Here the end of the transition line that
    begins at $E_{\oldstylenums1\textsc{rsb}}^+$ does not match
    $E_{\oldstylenums1\textsc{rsb}}^-$ but terminates at higher energies.
    $E_{\oldstylenums1\textsc{rsb}}^-$ still corresponds with the lower bound.
    \textbf{Bottom:} $\lambda=0.69$. Here the end of the transition line that
    begins at $E_{\oldstylenums1\textsc{rsb}}^+$ terminates at lower energies
    than $E_{\oldstylenums1\textsc{rsb}}^-$, and therefore its terminus defines
    the lower bound.
  } \label{fig:order}
\end{figure}

Fig.~\ref{fig:energy_ranges} shows the range of energies where nontrivial
correlations are found between stationary points in several $3+s$ models as
$\lambda$ is varied. For models with smaller $s$, such correlations are found
only among saddles, with the boundary never dipping beneath the minimum energy
of saddles $E_\mathrm{min}$. Also, these models have a transition boundary that
smoothly connects $E_{\oldstylenums1\textsc{rsb}}^+$ and
$E_{\oldstylenums1\textsc{rsb}}^-$, so $E_{\oldstylenums1\textsc{rsb}}^-$
corresponds to the lower bound of \textsc{rsb} complexity. For large enough
$s$, the range passes into minima, which is excepted as these models have
nontrivial complexity of their ground states. This also seems to correspond
with the decoupling of the \textsc{rsb} solutions connected to
$E_{\oldstylenums1\textsc{rsb}}^+$ and $E_{\oldstylenums1\textsc{rsb}}^-$, with
the two phase boundaries no longer corresponding, as in Fig.~\ref{fig:order}. In
these cases, $E_{\oldstylenums1\textsc{rsb}}^-$ sometimes gives the lower
bound, but sometimes it is given by the termination of the phase boundary
extended from $E_{\oldstylenums1\textsc{rsb}}^+$.

There are implications for the emergence of \textsc{rsb} in equilibrium.
Consider a specific $H$ with
\begin{equation}
  H(\pmb\sigma)
  =\frac{\sqrt\lambda}{p!}\sum_{i_1\cdots i_p}J^{(p)}_{i_1\cdots i_p}\sigma_{i_1}\cdots\sigma_{i_p}
  +\frac{\sqrt{1-\lambda}}{s!}\sum_{i_1\cdots i_s}J^{(s)}_{i_1\cdots i_s}\sigma_{i_1}\cdots\sigma_{i_s}
\end{equation}
where the interaction tensors $J$ are drawn from zero-mean normal distributions
with $\overline{(J^{(p)})^2}=p!/2N^{p-1}$ and likewise for $J^{(s)}$. Functions $H$ defined this way have the covariance
property \eqref{eq:covariance} with $f(q)=\frac12\big(\lambda
q^p+(1-\lambda)q^s\big)$. With the $J$s drawn in this way and fixed for $p=3$
and $s=14$, we can vary $\lambda$, and according to Fig.~\ref{fig:phases} we
should see a transition in the type of order at the ground state. What causes
the change? Our analysis indicates that stationary points with the required
order \emph{already exist in the landscape} as unstable saddles for small
$\lambda$, then eventually stabilize into metastable minima and finally become
the lowest lying states. This is different from the picture of existing
uncorrelated low-lying states splitting apart into correlated clusters. Where
uncorrelated stationary points do appear to split apart, when $\lambda$ is
decreased from large values, is among saddles, not minima.

A similar analysis can be made for other mixed models, like the $2+s$, which
should see complexities with other forms of \textsc{rsb}. For instance, in
\cite{Kent-Dobias_2023_How} we show that the complexity transitions from
\textsc{rs} to full \textsc{rsb} (\textsc{frsb}) along the line
\begin{equation}
  \mu
  =-\frac{(f'+f''(0))u_f}{(2f-f')f'f''(0)^{1/2}}
  -\frac{f''-f'}{f'-2f}E
\end{equation}
which can only be realized when $f''(0)\neq0$, as in the $2+s$ models. For
$s>2$, this transition line \emph{always} intersects the extremal line
\eqref{eq:extremal.line}, and so \textsc{rsb} complexity will always be found
among some population of stationary points. However, it is likely that for much
of the parameter space the so-called one-full \textsc{rsb}
({\oldstylenums1\textsc{frsb}}), rather than \textsc{frsb}, is the correct solution, as it likely is for
large $s$ and certain $\lambda$ in the $3+s$ models studied here. Further work to find the conditions for
transitions of the complexity to {\oldstylenums1\textsc{frsb}} and {\oldstylenums2\textsc{frsb}} is necessary. For values
of $s$ where there is trivial \textsc{rsb} in the ground state, we
expect that the {\oldstylenums1\textsc{rsb}} complexity is correct.

What are the implications for dynamics? We find that nontrivial correlations
tend to exist among saddle points with the largest or smallest possible index at
a given energy density, which are quite atypical in the landscape. However,
these strangely correlated saddle points must descend to uncorrelated minima,
which raises questions about whether structure on the boundary of a basin of
attraction is influential to the dynamics that descends into that basin. These
saddles might act as early-time separatrices for descent trajectories of certain algorithms.  With
open problems in even the gradient decent dynamics on these models (itself attracted to an atypical subset of marginal minima), it
remains to be seen whether such structures could be influential
\cite{Folena_2020_Rethinking, Folena_2021_Gradient, Folena_2023_On}. This structure among saddles
cannot be the only influence, since it seems that the $3+4$ model is `safe'
from nontrivial \textsc{rsb} among saddles.

We have determined the conditions under which the complexity of the mixed $3+s$
spherical models has different quenched and annealed averages, as the result of
nontrivial correlations between stationary points. We saw that these conditions
can arise among certain populations of saddle points even when the model is
guaranteed to lack such correlations between equilibrium states, and exist for
saddle points at a wide range of energies. This suggests that studies making
complexity calculations cannot reliably use equilibrium behavior to defend
the annealed approximation. Our result has direct implications for the
geometry of these landscapes, and perhaps could be influential to certain
out-of-equilibrium dynamics.

\paragraph{Funding information}

JK-D is supported by a \textsc{DynSysMath} Specific Initiative of the
INFN.

\printbibliography

@article{Altieri_2021_Properties,
 author = {Altieri, Ada and Roy, Felix and Cammarota, Chiara and Biroli, Giulio},
 title = {Properties of Equilibria and Glassy Phases of the Random {Lotka}-{Volterra} Model with Demographic Noise},
 journal = {Physical Review Letters},
 publisher = {American Physical Society (APS)},
 year = {2021},
 month = {6},
 number = {25},
 volume = {126},
 pages = {258301},
 url = {https://doi.org/10.1103%2Fphysrevlett.126.258301},
 doi = {10.1103/physrevlett.126.258301}
}

@article{Auffinger_2022_The,
 author = {Auffinger, Antonio and Zhou, Yuxin},
 title = {The Spherical {$p+s$} Spin Glass At Zero Temperature},
 year = {2022},
 month = {9},
 url = {http://arxiv.org/abs/2209.03866v1},
 date = {2022-09-08T15:08:26Z},
 eprint = {2209.03866v1},
 eprintclass = {math.PR},
 eprinttype = {arxiv}
}

@article{BenArous_2019_Geometry,
 author = {Ben Arous, Gérard and Subag, Eliran and Zeitouni, Ofer},
 title = {Geometry and Temperature Chaos in Mixed Spherical Spin Glasses at Low Temperature: The Perturbative Regime},
 journal = {Communications on Pure and Applied Mathematics},
 publisher = {Wiley},
 year = {2019},
 month = {11},
 number = {8},
 volume = {73},
 pages = {1732--1828},
 url = {https://doi.org/10.1002%2Fcpa.21875},
 doi = {10.1002/cpa.21875}
}

@article{Bray_2007_Statistics,
 author = {Bray, Alan J. and Dean, David S.},
 title = {Statistics of Critical Points of {Gaussian} Fields on Large-Dimensional Spaces},
 journal = {Physical Review Letters},
 publisher = {American Physical Society (APS)},
 year = {2007},
 month = {4},
 number = {15},
 volume = {98},
 pages = {150201},
 url = {https://doi.org/10.1103%2Fphysrevlett.98.150201},
 doi = {10.1103/physrevlett.98.150201}
}

@article{Castellani_2005_Spin-glass,
 author = {Castellani, Tommaso and Cavagna, Andrea},
 title = {Spin-glass theory for pedestrians},
 journal = {Journal of Statistical Mechanics: Theory and Experiment},
 publisher = {IOP Publishing},
 year = {2005},
 month = {5},
 number = {05},
 volume = {2005},
 pages = {P05012},
 url = {https://doi.org/10.1088%2F1742-5468%2F2005%2F05%2Fp05012},
 doi = {10.1088/1742-5468/2005/05/p05012}
}

@article{Cavagna_1998_Stationary,
 author = {Cavagna, Andrea and Giardina, Irene and Parisi, Giorgio},
 title = {Stationary points of the {Thouless}-{Anderson}-{Palmer} free energy},
 journal = {Physical Review B},
 publisher = {American Physical Society (APS)},
 year = {1998},
 month = {5},
 number = {18},
 volume = {57},
 pages = {11251--11257},
 url = {https://doi.org/10.1103%2Fphysrevb.57.11251},
 doi = {10.1103/physrevb.57.11251}
}

@article{Crisanti_1992_The,
 author = {Crisanti, A. and Sommers, H.-J.},
 title = {The spherical $p$-spin interaction spin glass model: the statics},
 journal = {Zeitschrift für Physik B Condensed Matter},
 publisher = {Springer Science and Business Media LLC},
 year = {1992},
 month = {10},
 number = {3},
 volume = {87},
 pages = {341--354},
 url = {https://doi.org/10.1007%2Fbf01309287},
 doi = {10.1007/bf01309287}
}

@article{Crisanti_2004_Spherical,
 author = {Crisanti, A. and Leuzzi, L.},
 title = {Spherical $2+p$ Spin-Glass Model: An Exactly Solvable Model for Glass to Spin-Glass Transition},
 journal = {Physical Review Letters},
 publisher = {American Physical Society (APS)},
 year = {2004},
 month = {11},
 number = {21},
 volume = {93},
 pages = {217203},
 url = {https://doi.org/10.1103%2Fphysrevlett.93.217203},
 doi = {10.1103/physrevlett.93.217203}
}

@article{Crisanti_2006_Spherical,
 author = {Crisanti, A. and Leuzzi, L.},
 title = {Spherical $2+p$ spin-glass model: An analytically solvable model with a glass-to-glass transition},
 journal = {Physical Review B},
 publisher = {American Physical Society (APS)},
 year = {2006},
 month = {1},
 number = {1},
 volume = {73},
 pages = {014412},
 url = {https://doi.org/10.1103%2Fphysrevb.73.014412},
 doi = {10.1103/physrevb.73.014412}
}

@article{Crisanti_2007_Amorphous-amorphous,
 author = {Crisanti, Andrea and Leuzzi, Luca},
 title = {Amorphous-amorphous transition and the two-step replica symmetry breaking phase},
 journal = {Physical Review B},
 publisher = {American Physical Society (APS)},
 year = {2007},
 month = {11},
 number = {18},
 volume = {76},
 pages = {184417},
 url = {https://doi.org/10.1103%2Fphysrevb.76.184417},
 doi = {10.1103/physrevb.76.184417}
}

@article{Crisanti_2011_Statistical,
 author = {Crisanti, A. and Leuzzi, L. and Paoluzzi, M.},
 title = {Statistical mechanical approach to secondary processes and structural relaxation in glasses and glass formers},
 journal = {The European Physical Journal E},
 publisher = {Springer Science and Business Media LLC},
 year = {2011},
 month = {9},
 number = {9},
 volume = {34},
 pages = {98},
 url = {https://doi.org/10.1140%2Fepje%2Fi2011-11098-3},
 doi = {10.1140/epje/i2011-11098-3}
}

@article{Folena_2020_Rethinking,
 author = {Folena, Giampaolo and Franz, Silvio and Ricci-Tersenghi, Federico},
 title = {Rethinking Mean-Field Glassy Dynamics and Its Relation with the Energy Landscape: The Surprising Case of the Spherical Mixed $p$-Spin Model},
 journal = {Physical Review X},
 publisher = {American Physical Society},
 year = {2020},
 month = {8},
 volume = {10},
 pages = {031045},
 url = {https://link.aps.org/doi/10.1103/PhysRevX.10.031045},
 doi = {10.1103/PhysRevX.10.031045},
 issue = {3},
 numpages = {26}
}

@article{Folena_2021_Gradient,
 author = {Folena, Giampaolo and Franz, Silvio and Ricci-Tersenghi, Federico},
 title = {Gradient descent dynamics in the mixed $p$-spin spherical model: finite-size simulations and comparison with mean-field integration},
 journal = {Journal of Statistical Mechanics: Theory and Experiment},
 publisher = {IOP Publishing},
 year = {2021},
 month = {3},
 number = {3},
 volume = {2021},
 pages = {033302},
 url = {https://doi.org/10.1088%2F1742-5468%2Fabe29f},
 doi = {10.1088/1742-5468/abe29f}
}

@article{Folena_2023_On,
 author = {Folena, Giampaolo and Zamponi, Francesco},
 title = {On weak ergodicity breaking in mean-field spin glasses},
 year = {2023},
 month = {2},
 url = {http://arxiv.org/abs/2303.00026v2},
 date = {2023-02-28T19:02:47Z},
 eprint = {2303.00026v2},
 eprintclass = {cond-mat.dis-nn},
 eprinttype = {arxiv}
}

@article{Fyodorov_2004_Complexity,
 author = {Fyodorov, Yan V.},
 title = {Complexity of Random Energy Landscapes, Glass Transition, and Absolute Value of the Spectral Determinant of Random Matrices},
 journal = {Physical Review Letters},
 publisher = {American Physical Society (APS)},
 year = {2004},
 month = {6},
 number = {24},
 volume = {92},
 pages = {240601},
 url = {https://doi.org/10.1103%2Fphysrevlett.92.240601},
 doi = {10.1103/physrevlett.92.240601}
}

@article{Fyodorov_2007_Density,
 author = {Fyodorov, Y. V. and Sommers, H.-J. and Williams, I.},
 title = {Density of stationary points in a high dimensional random energy landscape and the onset of glassy behavior},
 journal = {JETP Letters},
 publisher = {Pleiades Publishing Ltd},
 year = {2007},
 month = {5},
 number = {5},
 volume = {85},
 pages = {261--266},
 url = {https://doi.org/10.1134%2Fs0021364007050098},
 doi = {10.1134/s0021364007050098}
}

@article{Gershenzon_2023_On-Site,
 author = {Gershenzon, I. and Lacroix-A-Chez-Toine, B. and Raz, O. and Subag, E. and Zeitouni, O.},
 title = {On-Site Potential Creates Complexity in Systems with Disordered Coupling},
 journal = {Physical Review Letters},
 publisher = {American Physical Society (APS)},
 year = {2023},
 month = {6},
 number = {23},
 volume = {130},
 pages = {237103},
 url = {https://doi.org/10.1103%2Fphysrevlett.130.237103},
 doi = {10.1103/physrevlett.130.237103}
}

@article{Kent-Dobias_2021_Complex,
 author = {Kent-Dobias, Jaron and Kurchan, Jorge},
 title = {Complex complex landscapes},
 journal = {Physical Review Research},
 publisher = {American Physical Society (APS)},
 year = {2021},
 month = {4},
 number = {2},
 volume = {3},
 pages = {023064},
 url = {https://doi.org/10.1103%2Fphysrevresearch.3.023064},
 doi = {10.1103/physrevresearch.3.023064}
}

@article{Kent-Dobias_2023_How,
 author = {Kent-Dobias, Jaron and Kurchan, Jorge},
 title = {How to count in hierarchical landscapes: a full solution to mean-field complexity},
 journal = {Physical Review E},
 publisher = {American Physical Society (APS)},
 year = {2023},
 month = {6},
 number = {6},
 volume = {107},
 pages = {064111},
 url = {https://doi.org/10.1103/PhysRevE.107.064111},
 doi = {10.1103/PhysRevE.107.064111}
}

@article{Krakoviack_2007_Comment,
 author = {Krakoviack, V.},
 title = {Comment on ``Spherical {$2+p$} spin-glass model: An analytically solvable model with a glass-to-glass transition''},
 journal = {Physical Review B},
 publisher = {American Physical Society (APS)},
 year = {2007},
 month = {10},
 number = {13},
 volume = {76},
 pages = {136401},
 url = {https://doi.org/10.1103%2Fphysrevb.76.136401},
 doi = {10.1103/physrevb.76.136401}
}

@article{Krzakala_2007_Landscape,
 author = {Krzakala, Florent and Kurchan, Jorge},
 title = {Landscape analysis of constraint satisfaction problems},
 journal = {Physical Review E},
 publisher = {American Physical Society (APS)},
 year = {2007},
 month = {8},
 number = {2},
 volume = {76},
 pages = {021122},
 url = {https://doi.org/10.1103%2Fphysreve.76.021122},
 doi = {10.1103/physreve.76.021122}
}

@article{Muller_2006_Marginal,
 author = {Müller, Markus and Leuzzi, Luca and Crisanti, Andrea},
 title = {Marginal states in mean-field glasses},
 journal = {Physical Review B},
 publisher = {American Physical Society (APS)},
 year = {2006},
 month = {10},
 number = {13},
 volume = {74},
 pages = {134431},
 url = {https://doi.org/10.1103%2Fphysrevb.74.134431},
 doi = {10.1103/physrevb.74.134431}
}

@article{Ros_2019_Complex,
 author = {Ros, Valentina and Ben Arous, Gérard and Biroli, Giulio and Cammarota, Chiara},
 title = {Complex Energy Landscapes in Spiked-Tensor and Simple Glassy Models: Ruggedness, Arrangements of Local Minima, and Phase Transitions},
 journal = {Physical Review X},
 publisher = {American Physical Society (APS)},
 year = {2019},
 month = {1},
 number = {1},
 volume = {9},
 pages = {011003},
 url = {https://doi.org/10.1103%2Fphysrevx.9.011003},
 doi = {10.1103/physrevx.9.011003}
}

@article{Ros_2023_Generalized,
 author = {Ros, Valentina and Roy, Felix and Biroli, Giulio and Bunin, Guy and Turner, Ari M.},
 title = {Generalized {Lotka}-{Volterra} Equations with Random, Nonreciprocal Interactions: The Typical Number of Equilibria},
 journal = {Physical Review Letters},
 publisher = {American Physical Society},
 year = {2023},
 month = {6},
 volume = {130},
 pages = {257401},
 url = {https://link.aps.org/doi/10.1103/PhysRevLett.130.257401},
 doi = {10.1103/PhysRevLett.130.257401},
 issue = {25}
}

@article{Ros_2023_Quenched,
 author = {Ros, Valentina and Roy, Felix and Biroli, Giulio and Bunin, Guy},
 title = {Quenched complexity of equilibria for asymmetric Generalized {Lotka}-{Volterra} equations},
 year = {2023},
 month = {4},
 url = {http://arxiv.org/abs/2304.05284v1},
 date = {2023-04-11T15:34:53Z},
 eprint = {2304.05284v1},
 eprintclass = {cond-mat.dis-nn},
 eprinttype = {arxiv}
}

@article{Stein_1995_Broken,
 author = {Stein, D. L. and Newman, C. M.},
 title = {Broken ergodicity and the geometry of rugged landscapes},
 journal = {Physical Review E},
 publisher = {American Physical Society (APS)},
 year = {1995},
 month = {6},
 number = {6},
 volume = {51},
 pages = {5228--5238},
 url = {https://doi.org/10.1103%2Fphysreve.51.5228},
 doi = {10.1103/physreve.51.5228}
}

@article{Wainrib_2013_Topological,
 author = {Wainrib, Gilles and Touboul, Jonathan},
 title = {Topological and Dynamical Complexity of Random Neural Networks},
 journal = {Physical Review Letters},
 publisher = {American Physical Society (APS)},
 year = {2013},
 month = {3},
 number = {11},
 volume = {110},
 pages = {118101},
 url = {https://doi.org/10.1103%2Fphysrevlett.110.118101},
 doi = {10.1103/physrevlett.110.118101}
}

@article{Yang_2023_Stochastic,
 author = {Yang, Ning and Tang, Chao and Tu, Yuhai},
 title = {Stochastic Gradient Descent Introduces an Effective Landscape-Dependent Regularization Favoring Flat Solutions},
 journal = {Physical Review Letters},
 publisher = {American Physical Society (APS)},
 year = {2023},
 month = {6},
 number = {23},
 volume = {130},
 pages = {237101},
 url = {https://doi.org/10.1103%2Fphysrevlett.130.237101},
 doi = {10.1103/physrevlett.130.237101}
}

@unpublished{ElAlaoui_2020_Algorithmic,
 author = {El Alaoui, Ahmed and Montanari, Andrea},
 title = {Algorithmic Thresholds in Mean Field Spin Glasses},
 year = {2020},
 month = {9},
 url = {http://arxiv.org/abs/2009.11481v1},
 archiveprefix = {arXiv},
 date = {2020-09-24T04:22:42Z},
 eprint = {2009.11481v1},
 eprintclass = {cond-mat.stat-mech},
 eprinttype = {arxiv},
 primaryclass = {cond-mat.stat-mech}
}

@article{Subag_2020_Following,
 author = {Subag, Eliran},
 title = {Following the Ground States of Full-{RSB} Spherical Spin Glasses},
 journal = {Communications on Pure and Applied Mathematics},
 publisher = {Wiley},
 year = {2020},
 month = {6},
 number = {5},
 volume = {74},
 pages = {1021--1044},
 url = {https://doi.org/10.1002%2Fcpa.21922},
 doi = {10.1002/cpa.21922}
}

\end{document}